\documentclass[
    ,final            
  ]
 {aipproc}

\layoutstyle{6x9}

\usepackage{amssymb}
\newcommand{\be}{\begin{equation}}
\newcommand{\ee}{\end{equation}}
\newcommand{\bea}{\begin{eqnarray}}
\newcommand{\eea}{\end{eqnarray}}
\newcommand{\mpl}{M_{\rm pl}}

\begin{document}

\title{Non-linear non-local Cosmology}


\author{N. J. Nunes}{
  address={Department of Applied Mathematics and Theoretical Physics, Wilberforce Road, Cambridge, CB3 0WA, UK}
}

\author{D. J. Mulryne}{
  address={Department of Applied Mathematics and Theoretical Physics, Wilberforce Road, Cambridge, CB3 0WA, UK}
}


\begin{abstract}
Non-local equations of motion contain an infinite number of
derivatives and commonly appear in a number of string theory models.
We review how these equations can be rewritten in the form of a diffusion-like equation with non-linear boundary conditions.  Moreover, we show that this equation can be solved as an initial value problem once a set of non-trivial initial conditions that satisfy the boundary conditions is found. We find these initial conditions by looking at the linear approximation to the boundary conditions. We then numerically solve the diffusion-like equation,  and hence the non-local equations, as an initial value problem for the full non-linear potential and subsequently identify the cases when inflation is attained.
\end{abstract}

\maketitle


\section{Introduction}

Several string theory models such as the $p$-adic string and the cubic  (super-)string field theory (CSSFT) \cite{background} lead to non-local field equations of motion 
that involve an infinite number of derivatives.
These models give rise to interesting dynamics and have been considered in the context of cosmology \cite{cosmology}, particularly inflation as well as bouncing universes. 
In general, we can write their action as follows:
\begin{equation}
\label{action}
S=\int d^4x \sqrt{-g} \left [ \frac{\mpl^2}{2} R + \gamma^4 \lambda \left
  (\frac{1}{2} \psi F(\Box) \psi - V(\psi) \right ) \right ] \,,
\end{equation}
where the d'Alembertian operator is
\begin{equation}
\Box \equiv \frac{1}{\sqrt{-g}} \partial_\mu \left( \sqrt{-g} \, g^{\mu
  \nu}\partial_\nu \right) = -\frac{d^2}{dt^2} - 3H \frac{d}{dt} \,,
\end{equation}
where the latter equality only applies for a Friedmann-Robertson-Walker universe. $V(\psi)$ is the scalar potential and $F(\Box) = -(1+4\xi^2\alpha \Box) e^{-\alpha \Box}$. The field equations are therefore 
\be
\label{eom}
(1+4\xi^2\alpha \Box) e^{-\alpha \Box} \psi= -\frac{d V(\psi)}{d \psi} \,.
\ee

As it stands, the computation of the evolution of $\psi(t)$ seems to be a hopeless task as it involves an infinite number of derivatives. Indeed, for a general differential equation of the form
\be
\frac{d^n \psi}{d t^n} = f(t,\psi,\dot{\psi},...,\psi^{(n-1)}) \,,
\ee
with finite $n$, we must specify initial conditions $\psi_i$, $\dot{\psi}_i$, ..., $\psi^{(n-1)}_i$. We may naively think that as $n \rightarrow \infty$ any solution is allowed, as we would be able to write $\psi(t)$ as a Taylor expansion whose coefficients can arbitrarily be chosen. While this is not the case, as the field equation itself imposes an infinite number of constraints on the allowed initial conditions \cite{moeller}, formulating 
an initial value problem (IVP) for the field's evolution still seems hopeless. This leads us to question whether we can reinterpret the problem in a helpful manner.

Defining 
$\Psi(t,r) \equiv e^{-r \alpha \Box}\psi(t)$, where $r$ is an auxiliary variable, and differentiating with respect to $r$ we obtain a diffusion-like equation \cite{diffusion}
\be
\label{diffusioneq}
\Box\Psi(t,r)=-\frac{1}{\alpha} \, \frac{\partial \Psi(t,r)}{\partial r} \,.
\ee
Given that $\Psi(t,1) = e^{-\alpha\Box}\psi(t)$ and $\Psi(t,0) = \psi(t)$, the equation of motion (\ref{eom}) leads to the non-trivial boundary conditions
\be
\label{boundary}
\Psi(t,1) - 4\xi^2 \left[\frac{\partial \Psi(t,r)}{\partial r}\right]_{r=1} =
-\frac{\partial V(\Psi(t,0))}{\partial \Psi(t,0)} \,.
\ee

 In principle, the pde (\ref{diffusioneq}) can now be solved as an IVP by specifying an initial profile $\Psi(t_{\rm i},r)$, which satisfies the boundary conditions (\ref{boundary}), and evolving it forwards in time. For the diffusion-like pde (\ref{diffusioneq}) this is, however, a numerically unstable procedure. In this work we will 
solve (\ref{diffusioneq}) as an IVP, but the instability requires us to employ the regularisation methods described in Ref.~\cite{Mulryne:2008iq}.

Finding the initial profile of $\Psi(t_{\rm i},r)$ is non-trivial. As an example let us take $\xi^2 = 0$ and $V(\psi) = -\psi^4/4$, which is the case of the $p$-adic string, and also assume that $\dot{\Psi}(t_{\rm i},r) = 0$. Applying the d'Alembertian to the boundary condition and using the diffusion equation we find an infinite set of conditions that the initial profile $\Psi(t_{\rm i},r)$ must satisfy. 
\bea
\Psi(t_{\rm i},1) &=& \Psi(t_{\rm i},0)^3  \,, \\ 
\left[ \frac{\partial \Psi(t_{\rm i},r)}{\partial r}\right]_{r=1} &=& 
3 \,\Psi(t_{\rm i},0)^2  \left[\frac{ \partial \Psi(t_{\rm i},r)}{\partial r}\right]_{r=0}  \,, 
\eea
etc. This is equivalent to having to specify all the constraints on the allowed initial conditions that we mentioned before. It seems that the diffusion-like equation failed to bring us any advantage. In the next section, however, we are going to see that this infinite set of conditions is trivially satisfied when the equations are linearised which gives us hope to be able to go beyond the linear regime using the diffusion-like equation.

\section{Linearised equations}

By linearising $\psi$ around a value $A$ such that $\psi(t) = A + \phi(t)$ and expanding the action (\ref{action}) to second order, we obtain the new field equations 
\be
(1+4\xi^2\alpha \Box) e^{-\alpha\Box} \phi = - \frac{dU(\phi)}{d\phi} \,,
\ee
where $U(\phi) = U_0 - c \, \phi - \frac{1}{2} m^2 \phi^2$ and the coefficients are related to the original scalar potential $V(\psi)$: 
$U_0 = V(A)+A^2/2$, $c = -V'(A)-A$,  $m^2 = -V''(A)$. As we have done before, we can now define $\Phi(t,r) = e^{-r \alpha \Box} \phi(t)$ from which follows the new diffusion-like equation
\be
\label{diffusioneq2}
\Box\Phi(t,r)=-\frac{1}{\alpha} \, \frac{\partial \Phi(t,r)}{\partial r} \,,
\ee
with the linear boundary condition
\be
\Phi(t,1) - 4 \xi^2 \left[\frac{\partial \Phi(t,r)}{\partial r}\right]_{r = 1} = m^2 \Phi(t,0) + c \,.
\ee
Going back to our $p$-adic string example with $\xi^2 = 0$, $V(\psi) = -\psi^4/4$, and using $A = 1$, applying $\Box$ successively to the boundary condition and using the diffusion-like equation we have the infinite set of conditions on the initial profile:
\bea
\Phi(t_{\rm i},1) &=& 3 \, \Phi(t_{\rm i},0)   \\ 
\left[\frac{\partial \Phi(t_{\rm i},r)}{\partial r}\right]_{r=1} &=& 3 \left[\frac{ \partial
      \Phi(t_{\rm i},r)}{\partial r}\right]_{r=0}  \,, 
\eea
etc, and clearly, $\Phi(t_{\rm i},r) = \exp(r \, \ln 3)$ satisfies all of these. 

For a general model and potential, employing separation of variables, the solution for $\Phi(t,r)$ is
\be
\Phi(t,r) = \phi(t) e^{\alpha\omega^2 \, r} + \frac{b}{\omega^2}\left(e^{\alpha\omega^2 \, r} -1\right) \,,
\ee
where $\phi(t)$ solves a local equation of motion
%
\be
\Box\phi = -\omega^2 \phi - b \,,
\ee
where $b = c \omega^2/(m^2-1)$ 
and $\omega^2$ satisfies the characteristic equation: 
\be
e^{\alpha\omega^2}(1-4\xi^2\alpha\omega^2) = m^2 \,.
\ee
This equation admits in general more than one root which can be complex,
depending on the value of $m^2$ and whether $\xi^2$ vanishes or not. General conclusions can be drawn from the reality  of $\omega^2$ alone. In particular, if a root $\omega^2$ is real, so must be $\phi$ and we can define 
$\varphi = B \, \phi$ and  $\tilde{b} = B\, b$,  where
$B^2 = \gamma^4 \lambda \alpha \left[1-4\xi^2(1+4\alpha\omega^2)\right] e^{\alpha\omega^2}$ such that $\varphi$ is a canonically normalised field satisfying the local equation $\Box\varphi =  -\omega^2 \varphi - \tilde{b}$.
On the other hand, if $\omega^2$ is complex, its complex conjugate is also a root and we must search for solutions of the kind $\phi = \sum_m \phi_m$
such that $\phi$ is real even though the various $\phi_m$ are not. Redefining the fields $\phi_m$ in terms of $\varphi_m$ as before, and further expressing 
these in terms of the canonically normalised real fields $\chi_m$ and 
$\sigma_m$, $\varphi_m = (\chi_m + i \sigma_m)/\sqrt{2}$, 
$\varphi_m^* = (\chi_m - i \sigma_m)/\sqrt{2}$ , we obtain their respective equations of motion
\bea
\Box\chi_m &=&  
-\alpha_m^2 \chi_m + \beta_m^2 \sigma_m -\tilde{p}_m \,. \\
\Box \sigma_m &=& -\beta_m^2 \chi_m -\alpha_m^2\sigma_m - \tilde{q}_m \,.
\eea
This system of equations describes what is called a "quintom". A set of two real fields where one is a standard field and the other is a ghost.

\section{The $p$-adic string}
\begin{figure}
  \includegraphics[height=.25\textheight]{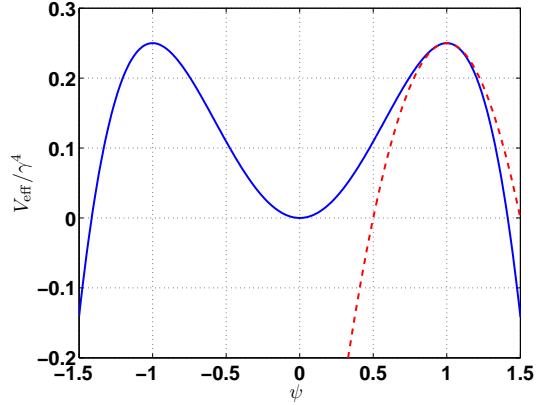}
  \caption{\label{veff1} The  $p$-adic string  effective potential  with $p = 3$. The solid and dashed lines represent the full potential and the quadratically expanded potential around the hilltop, respectively.}
\end{figure}

We will now look at two examples of the evolution of $\psi(t)$. We will start with the $p$-adic string for which we have $\gamma^4= m_s^4 p^2/g_s^2(p-1)$, $\lambda=1$, $V(\psi)=-\psi^{p+1}/(p+1)$, 
$\xi^2=0$, and $\alpha = \ln( p) /m_s^2$, hence, the characteristic equation is simply  $e^{\alpha\omega^2} = m^2$,  with $m^2 = p A^{p-1}$ and provided $A \neq 0$, the solutions are $ \alpha \omega^2 = \ln m^2  \pm 2n \pi \, i$, i.e., it admits one real solution and an infinite number of quintoms which imply that an infinite number of initial conditions must be set, in agreement with Ref.~\cite{ivp}. In this work we will assume than only some of the degrees of freedom are dynamical and consider separately the evolution of $\Psi(t,r)$ for individual values of $n$.
The effective potential $V_{\rm eff}(\psi) = \gamma^4(V(\psi) + \psi^2/2)$ is illustrated in Fig.~\ref{veff1}.

\begin{figure}
  \includegraphics[height=.27\textheight]{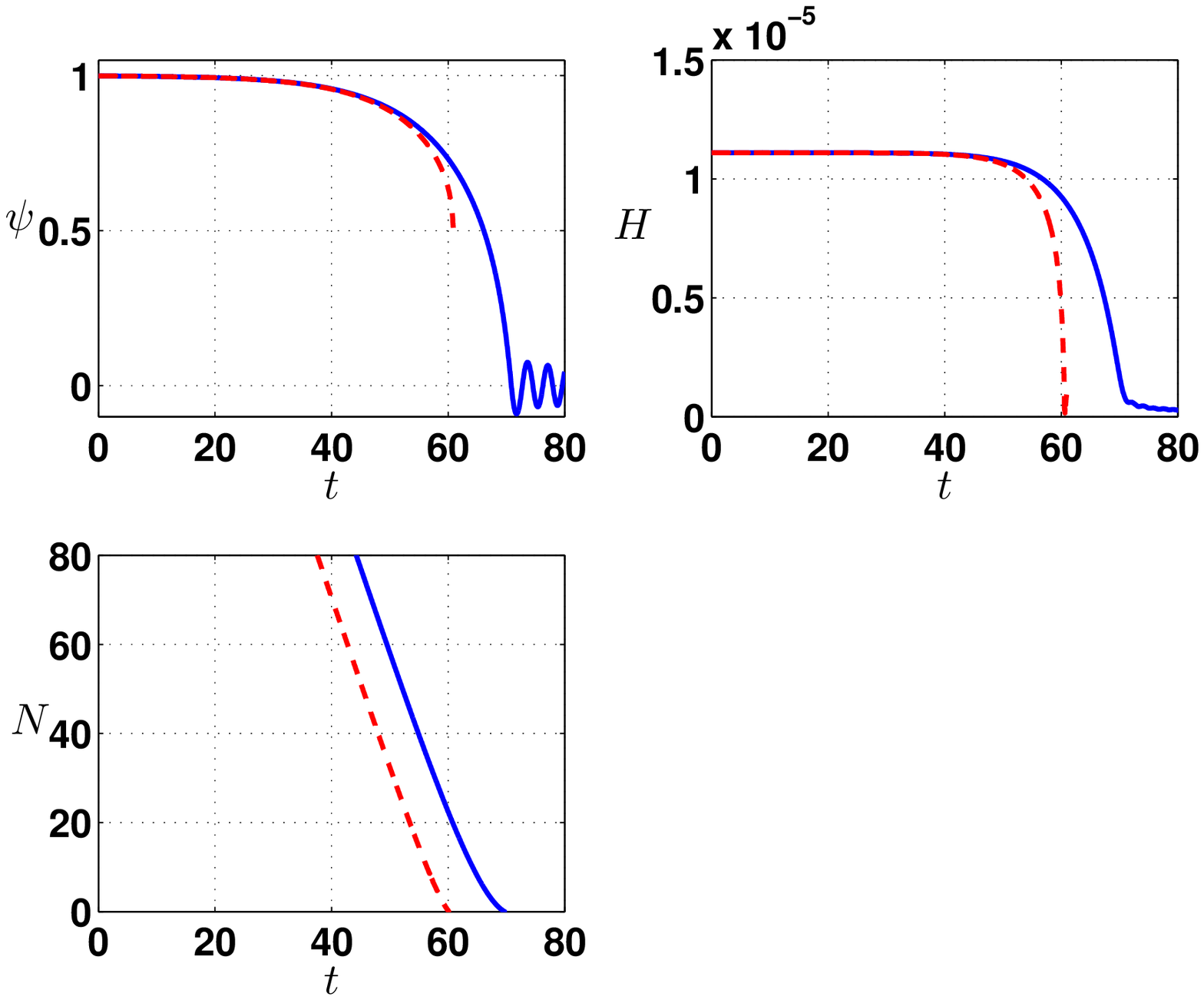}
  \includegraphics[height=.27\textheight]{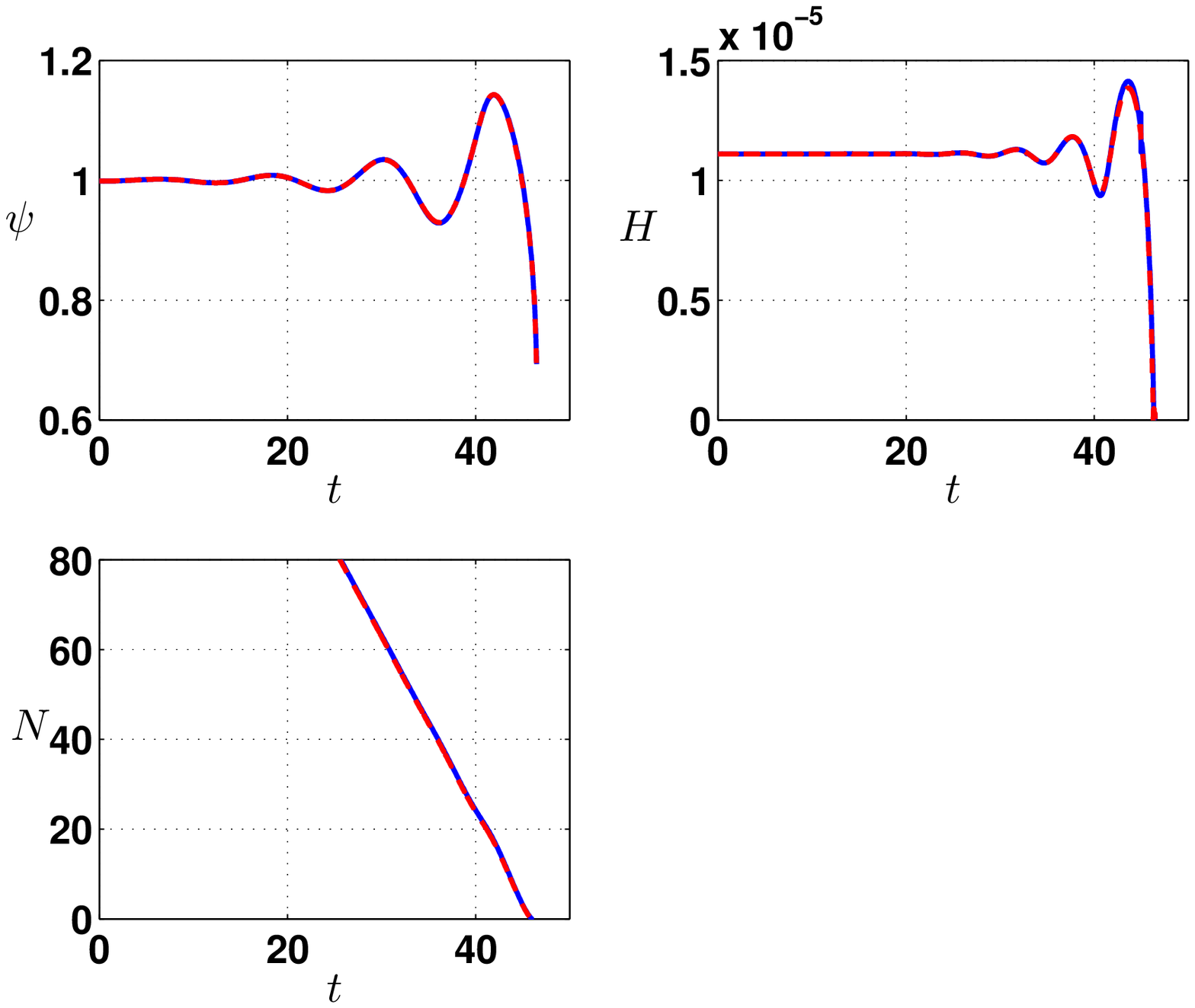}
  \caption{\label{evol1} The left panel shows the evolution of the field $\psi$, the Hubble rate and the number of $e$-folds of inflation in the $p$-adic string with $p=3$ and $n=0$. The right panel shows the same quantities for the first quintom, $n=1$. Solid and dashed lines represent the full non-linear and linear evolutions, respectively.}
\end{figure}

In Fig.~\ref{evol1} we show the evolution of $\psi$ with initial conditions very near the hilltop. For the real field ($n=0$), 
$\Psi(t_{\rm i},r) = 1 - \epsilon \, e^{\alpha\omega^2_R \, r}$ and  
$\dot{\Psi}(t_{\rm i})=0$ with $\epsilon = 10^{-3}$, $\psi$ decays to the minimum of the effective potential and oscillates . We also show the evolution of the Hubble rate and compute the number of $e$-folds of inflation, which can be seen to be sufficiently large to solve the horizon problem. For the first quintom ($n=1$), 
$\Psi(t_{\rm i},r) = 1 - \epsilon \, e^{\alpha\omega^2_R \, r} \, \cos(\alpha\omega^2_I r)$, and
$\dot{\Psi}(t_{\rm i},r) = 0$, 
we see that the field performs oscillations around the hilltop that grow in time and are propagated to the evolution of the Hubble rate. Eventually the amplitude of the Hubble rate is large enough to cause this to 
vanish and a recollapse occurs. We also show the evolution of the field had we considered only the quadratic approximation of the potential at the hilltop.

\section{The CSSFT string}
\begin{figure}
  \includegraphics[height=.25\textheight]{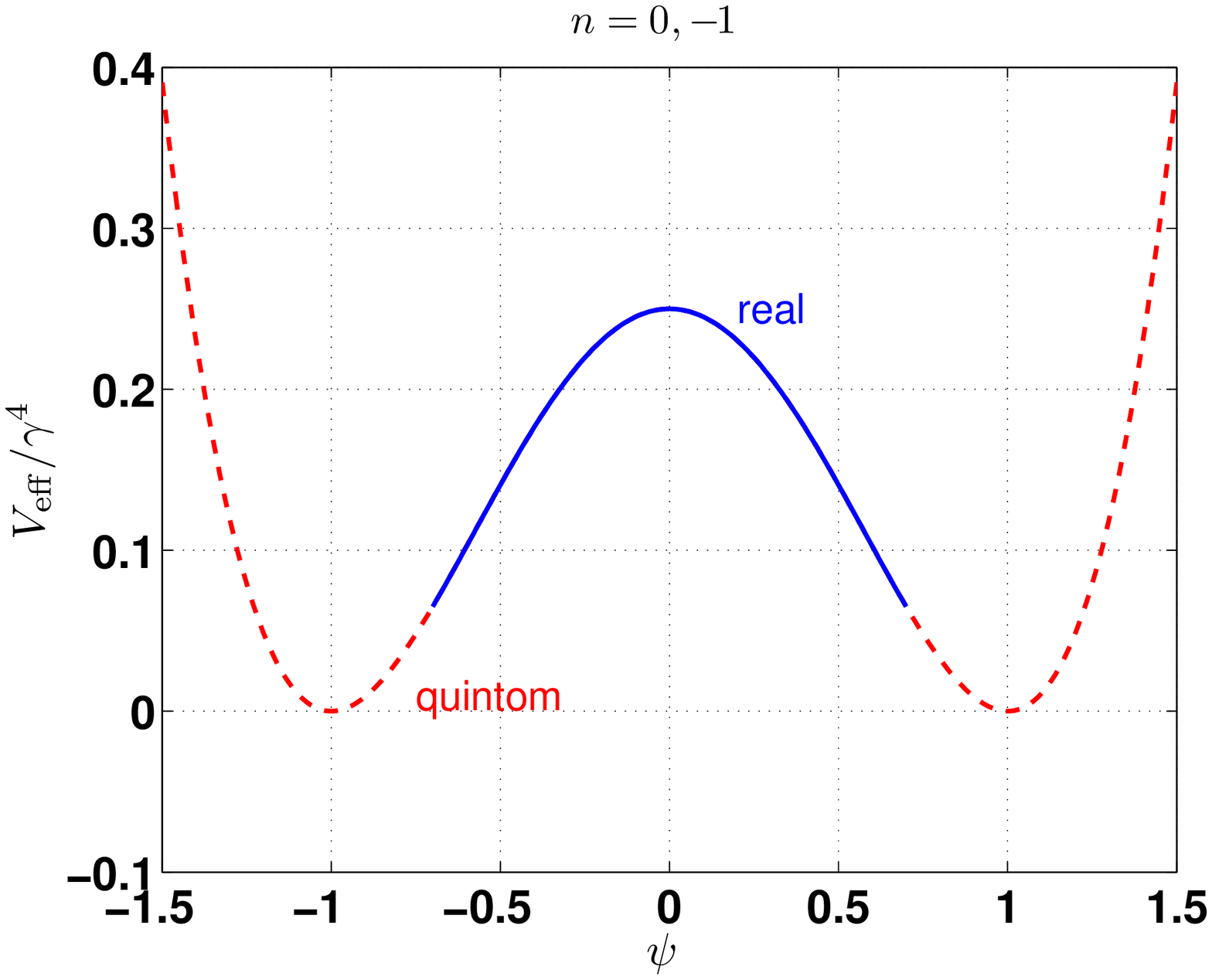}
  \includegraphics[height=.25\textheight]{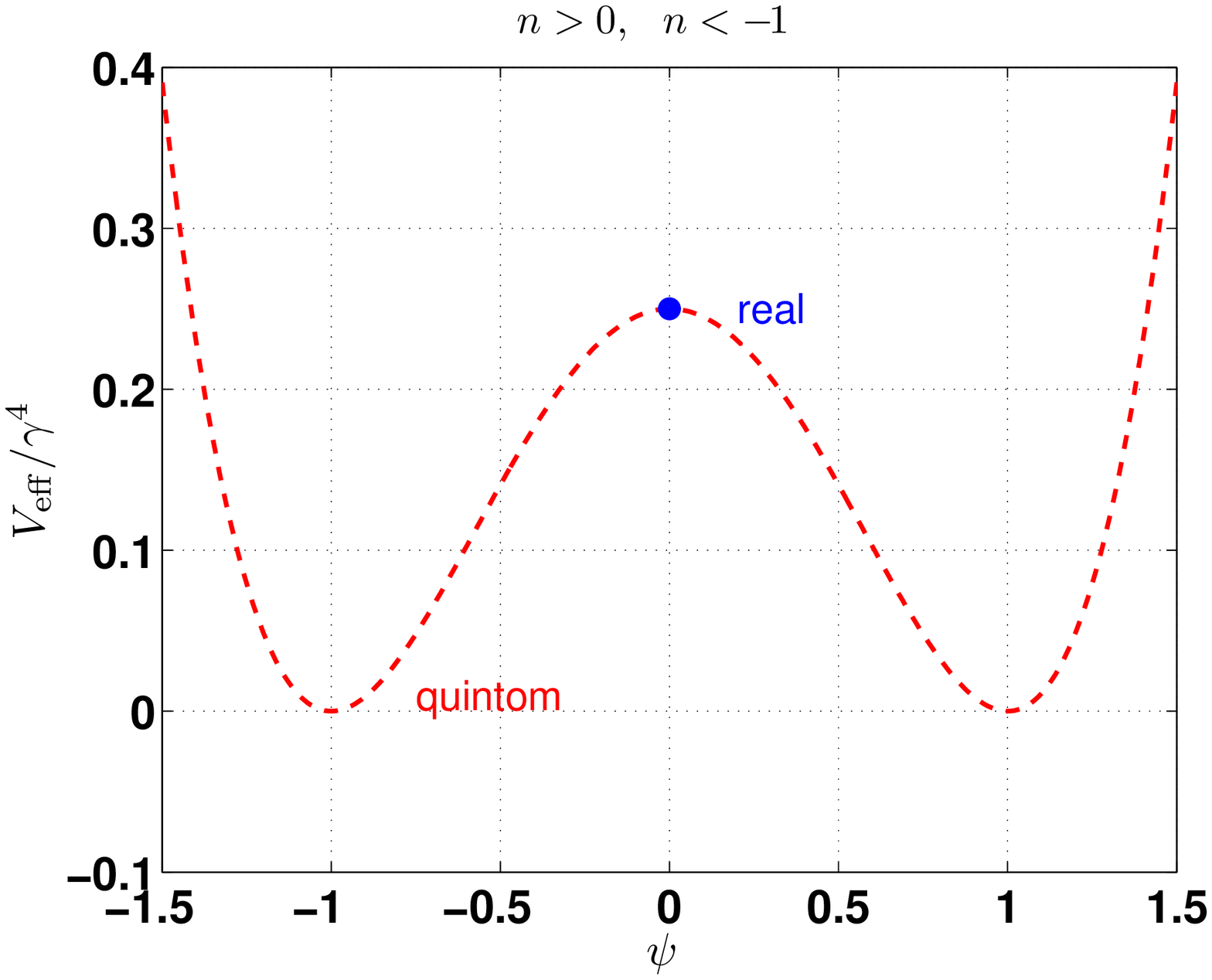}
  \caption{\label{pot2} The effective potential of the CSSFT string. The left panel shows the regions where the dynamics is described as single real field (solid line) or as quintom (dashed lines), when $n=0,-1$. The right panel shows the same regions for the remaining values of $n$.}
\end{figure}

In this section we will look at the dynamics of the field in the cubic superstring field theory. Here we have 
$\gamma^4= m_s^4/g_s^2$,
$\lambda=-1$, 
$V(\psi)=\psi^4/4$, 
$\xi^2=0.96$, 
$\alpha = 1/4 m_s^2$  with characteristic equation
$e^{\alpha\omega^2}(1-4\xi^2\alpha\omega^2) = m^2$, and
$m^2 = p A^{p-1}$. In this case, if $A\neq 0$, the roots of the characteristic equation are of the form $\alpha \omega^2 = 1/4\xi^2 + 
{\rm LambertW}\left(n,x \right)$ where $x = -m^2 e^{-1/4\xi^2}/4\xi^2$. If $A = 0$, then $\alpha \omega^2 = 1/4\xi^2$. In Fig.~\ref{pot2} we show the effective potential of this model and indicate the values of $\psi$ where the roots are real and where they are complex (and therefore the dynamics is described by quintoms). It is clear that in this case the minimum of the potential only admits quintoms.

\begin{figure}
  \includegraphics[height=.27\textheight]{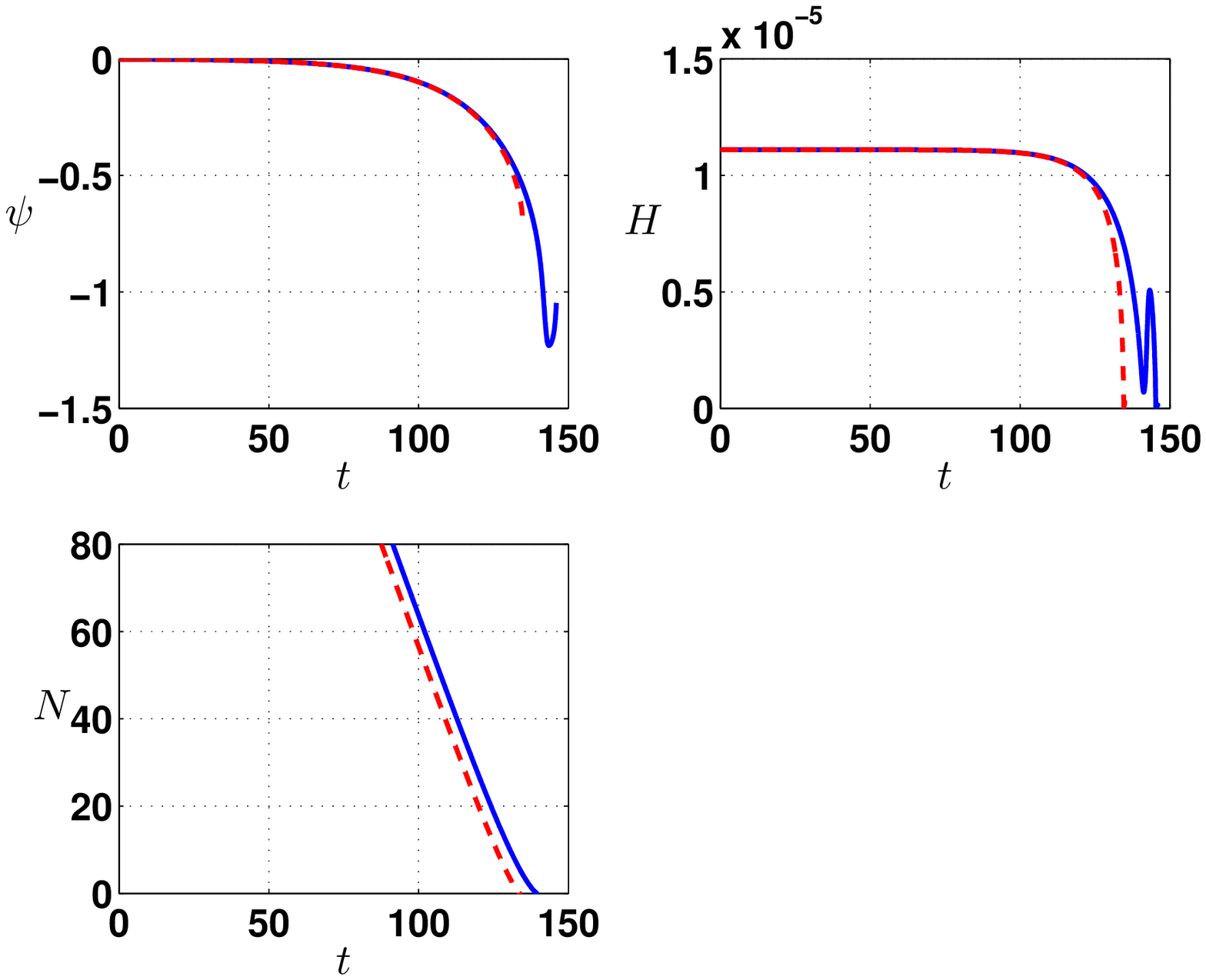}
  \includegraphics[height=.27\textheight]{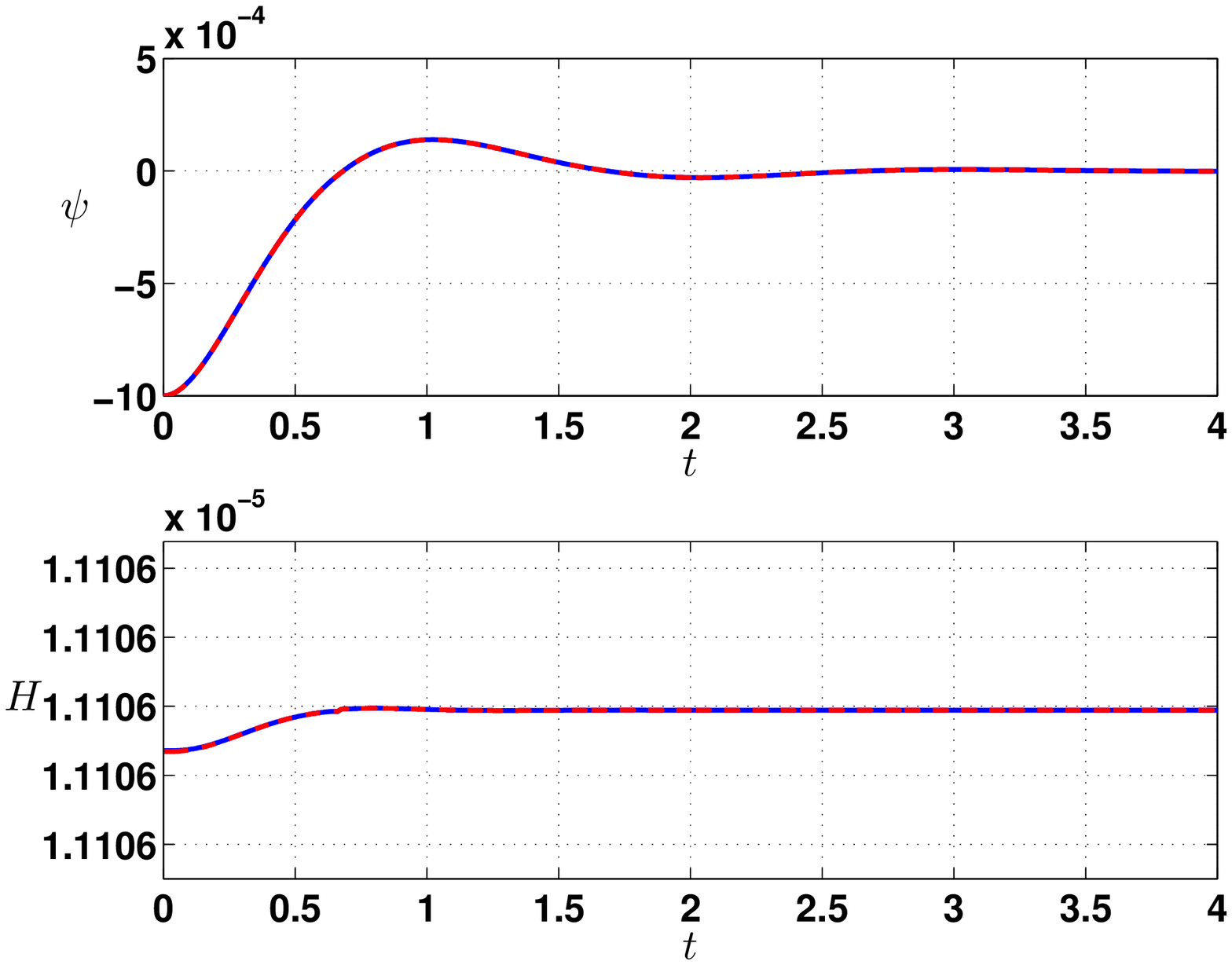}
  \caption{\label{evol2} The left panel shows the evolution of the field $\psi$, the Hubble rate and the number of $e$-folds of inflation in the CSSFT string with $n=0$. The right panel shows the same quantities for $n=1$. Solid and dashed lines represent the full non-linear and linear evolutions, respectively.}
\end{figure}

The evolution of the field with $n=0$ is shown in Fig.~\ref{evol2} for initial conditions 
$\Psi(t_{\rm i},r) = \epsilon \, e^{\alpha\omega^2_R \, r}$ and 
$\dot{\Psi}(t_{\rm i},r) = 0$. This corresponds to a real field evolving initially. We see that at late times, unlike in  the $p$-adic string, the Hubble rate does go to zero indicating that a quintom has appeared. This is expected as only quintoms are expected at the minimum of the effective potential. For the $n=1$ case, the evolution is also distinct from what happens in the $p$-adic string. Here the quintom decays towards the maximum of the potential and inflation lasts forever.


\begin{theacknowledgments}
  NJN thanks the organisation of "The dark side of the Universe" for a stimulating workshop.
\end{theacknowledgments}

\bibliographystyle{aipproc}   

\bibliography{sample}

\end{document}